\documentclass[10pt,conference]{IEEEtran}
\IEEEoverridecommandlockouts
\usepackage{framed}

\usepackage{cite}
\usepackage{graphicx}
\usepackage{textcomp}
\usepackage{soul}
\usepackage{xcolor}
\usepackage{subfigure}
\usepackage{listings}
\usepackage{url}
\usepackage{balance}
\usepackage{xspace}
\usepackage{bbding}
\usepackage{longtable}
\usepackage{booktabs}
\usepackage{array}
\usepackage{paralist}
\usepackage{multirow,makecell}
\usepackage{enumerate}
\usepackage{fancynum}
\usepackage{times}
\usepackage{fancyhdr,graphicx,amsmath,amssymb}
\usepackage[ruled,linesnumbered,boxed]{algorithm2e}
\usepackage{diagbox}

\usepackage[most]{tcolorbox}

\usepackage[hidelinks]{hyperref}
\usepackage{adjustbox}

\usepackage{flushend}

\definecolor{codegreen}{rgb}{0,0.6,0}
\definecolor{codegray}{rgb}{0.5,0.5,0.5}
\definecolor{codepurple}{rgb}{0.58,0,0.82}
\definecolor{backcolour}{rgb}{0.95,0.95,0.92}

\definecolor{KWColor}{rgb}{0,0,0}
\definecolor{CommentColor}{rgb}{0.4,0.4,0.4}
\definecolor{StringColor}{rgb}{0.4,0.4,0.4}
\newboolean{showcomments}
\setboolean{showcomments}{true}
\ifthenelse{\boolean{showcomments}}
{ \newcommand{\mynote}[2]{
    \fbox{\bfseries\sffamily\scriptsize#1}
    {\small$\blacktriangleright$\textsf{\emph{#2}}$\blacktriangleleft$}}}
{ \newcommand{\mynote}[2]{}}

\newcommand{\ie}{i.e.,\xspace}
\newcommand{\eg}{e.g.,\xspace}

\newcommand\tool[1]{\textsc{AndroSea}}

\pdfoutput=1
\usepackage{hyperref}
\hypersetup{
  pdfinfo={
    Title={title},
    Author={author},
  }
}
\usepackage{pdfpages}

\begin{document}

\title{Identifying and Characterizing Silently-Evolved Methods in the Android API}

\author{\IEEEauthorblockN{Pei Liu\IEEEauthorrefmark{1},
Li Li\IEEEauthorrefmark{1}$^{\alpha}$\thanks{$^{\alpha}$Corresponding author},
Yichun Yan\IEEEauthorrefmark{2},
Mattia Fazzini\IEEEauthorrefmark{2},
John Grundy\IEEEauthorrefmark{1}}
\IEEEauthorblockA{\IEEEauthorrefmark{1}Monash University, Melbourne, Australia \\
\{pei.liu,li.li,john.grundy\}@monash.edu}
\IEEEauthorblockA{\IEEEauthorrefmark{2}University of Minnesota, Minneapolis, USA \\
\{yan00104,mfazzini\}@umn.edu}}

\maketitle

\begin{abstract}
With over 500,000 commits and more than 700 contributors, the Android platform is undoubtedly one of the largest industrial-scale software projects. This project provides the Android API, and developers heavily rely on this API to develop their Android apps.
Unfortunately, because the Android platform and its API evolve at an extremely rapid pace, app developers need to continually monitor API changes to avoid compatibility issues in their apps (\ie issues that prevent apps from working as expected when running on newer versions of the API).
Despite a large number of studies on compatibility issues in the Android API, the research community has not yet investigated issues related to silently-evolved methods (SEMs).
These methods are functions whose behavior might have changed but the corresponding documentation did not change accordingly. Because app developers rely on the provided documentation to evolve their apps, changes to methods that are not suitably documented may lead to unexpected failures in the apps using these methods.

To shed light on this type of issue, we conducted a large-scale empirical study in which we identified and characterized SEMs across ten versions of the Android API. In the study, we identified SEMs, characterized the nature of the changes, and analyzed the impact of SEMs on a set of 1,000 real-world Android apps. Our experimental results show that SEMs do exist in the Android API, and that 957 of the apps we considered use at least one SEM.
Based on these results, we argue that the Android platform developers should take actions to avoid introducing SEMs, especially those involving semantic changes. This situation highlights the need for automated techniques and tools to help Android practitioners in this task.
\end{abstract}



\section{Introduction}

Mobile applications (or simply apps) are becoming increasingly prevalent in our lives. For instance, in 2017, US-based users spent an average of two hours and 25 minutes per day using mobile apps. This time accounts for more than 80\% of the total time spent by the users on their mobile devices~\cite{eMarketer2020}. Android apps, which run on the Android operating system (OS), are the most widely used type of mobile apps and account for over 70\% of the mobile OS market share~\cite{statista-1-2020}.
Android apps are not only extremely popular, but their number is also growing at a staggering speed, with about 35,000 new apps released on Google Play every month~\cite{statista-2-2020}.

One common trait of Android apps is that they rely heavily on the underlying Android OS, as the OS offers access to a large number of popular and essential app services. Apps can access these services using the application programming interface (API) of the OS. On average, 25\% of all methods and field references in the apps are uses of the Android API~\cite{mcdonnell2013empirical}. This characteristic facilitates app development~\cite{syer2015studying}, but it also creates a tight coupling between the apps and the version of the API used by the apps.

Unfortunately, the Android OS and its API evolve rapidly~\cite{mcdonnell2013empirical, linares2013api, bavota2014impact, yang2018how, li2018moonlightbox, li2018characterising}, and when a new version of the API is released, app developers need to carefully understand the changes to the API so that they can suitably adapt their apps to also run on the new version of the API. To help app developers in this task, Android provides a curated documentation that app developers can access using the platform website~\cite{google-api-2020}. This documentation is based on the comments associated with the source code of the API, and the comments are created and maintained by the API developers.

When API developers are working on a new version of the API, they not only should create comments that describe newly added API components but they should also update existing comments to report changes in existing API components (\eg~\cite{google-getAllNetworkInfo-2020}). In the latter task, developers should document both syntactic and behavioral changes introduced in the new version of the API. Although it is important to document both types of changes, it is imperative to document the second class of changes as app developers would otherwise not easily know how to update their apps, and users might experience field failures due to compatibility issues.

Although related work investigated a number of aspects associated with the evolution of the Android API~\cite{mcdonnell2013empirical, bavota2014impact, li2018cid, he2018understanding, li2020cda, li2016accessing}, to the best of our knowledge, no work has systematically analyzed whether there are issues in the documentation of the behavioral changes in the platform API. Additionally, no study identified the extent to which these issues might affect Android apps. A study on these topics would provide insights for building automated techniques that detect the issues and highlight mitigation strategies against the issues.


To fill this gap, we present an extensive empirical study that identifies and characterizes \textit{silently-evolved methods} (SEMs) in the Android API.
A SEM is a method whose implementation is different across two subsequent versions of the Android API, while the method's documentation (\ie the method's comment) is not changed. In the study, we (i) identify SEMs across different versions of the Android API, (ii) report the characteristics of these methods, and (iii) analyze the impact that SEMs might have on real-world apps. Specifically, we analyzed method updates across ten API releases and manually classified updates to methods whose documentation did not change. 
In the study, we identified 4,769 SEMs, which include 2,271 publicly-accessible methods.
After manually analyzing a statistically significant sample containing 562 SEMs, we found that 363 of the methods include semantic changes. Furthermore, we also analyzed the use of SEMs in a sample of 1,000 real-world Android apps and observed that 957 of these apps use at least one SEM. Interestingly, a number of such usages have been manually mitigated by app developers through API version checks\footnote{Version checks are recommended by Google to tackle API-induced compatibility issues. A typical check conforms to the following structure: if $\mathtt{SDK\_INT} < \mathit{n}$, call a method of the older API, otherwise, do something else. $\mathtt{SDK\_INT}$ is the short version for $\mathtt{android.os.Build.SDK\_INT}$ and this value, at runtime, provides the version of the API on which a certain app is running. This check ensures that the method of the API, which may introduce compatibility issues in the app when the app is running on newer versions of the API, will only be invoked if the app is running on API versions that are older than the one where the change was introduced.}, indicating that SEMs could indeed introduce compatibility issues in Android apps.
Overall, we believe that our results highlight that Android developers do not always thoroughly document semantic changes in the platform API and that these changes can extensively affect real-world Android apps.

In summary, the main contributions of this paper are:

\begin{itemize}
    \item A characterization of SEMs across the ten most used versions (at the time we started our study) of the Android API. We analyzed method updates across ten versions of the Android API and manually confirmed that 363 methods out of a sample of 562 publicly-accessible methods contain semantic changes. We also characterized the nature of SEMs and found that developers might have accidentally introduced the majority of them as most of them are evolved only once.

    \item A study of how SEMs impact Android apps. We investigated whether and how Android apps use SEMs in a sample of 1,000 real-world Android apps. We found that SEMs are commonly used by Android apps, and such usages could indeed lead to compatibility issues as app developers have added checks to prevent the execution of SEMs on certain versions of the Android API.
    
    \item A tool to identify SEMs and the experimental data containing the findings of our study. To identify and characterize SEMs, we designed a technique and implemented the approach in a prototype tool called \tool{}. \tool{} identifies methods across two versions of the Android API that have same signature, same method comment, but different method body. The tool and the experimental data are publicly available at \url{https://github.com/MobileSE/AndroSea}.
\end{itemize}

The remainder of this paper is organized as follows. Section~\ref{sec:terminology} defines relevant terminology. Section~\ref{sec:methodology} presents our study methodology. Section~\ref{sec:experiment} details the results of the study. We discuss implications for researchers and practitioners in Section~\ref{sec:discussion}. Section~\ref{sec:related} outlines related work. Finally, Section~\ref{sec:conclusion} provides concluding remarks.

\section{Terminology and Motivation}
\label{sec:terminology}

This sections introduces the relevant terminology we will use in the rest of this paper.
Consider two API versions (or levels): \textit{old API} = $\lbrack\mathit{m_{1}}$,
..., $\mathit{m_{k}}\rbrack$ and \textit{new API} =
$\lbrack\mathit{m_{1}^{\prime}}$, ...,
$\mathit{m_{l}^{\prime}}\rbrack$. 
A method $m_{i}$ from the old API is a
\textit{silently-evolved method} (SEM) if there is a method $m_{i}^{\prime}$
in the new API that shares the same signature and the same comment with $m_{i}$, but has a different
method body with respect to $m_{i}$.
Among SEMs, we define those methods that are publicly accessible (\ie methods that are declared as public) as
\textit{publicly-accessible silently-evolved methods} (PASEMs). In this paper, we focus on both SEMs and PASEMs as both classes might affect the execution of an android app.

Listing~\ref{lst:sample_snippet} provides an example of a PASEM, which shows the evolution of the method {\small \texttt{getSqlStatementType}} between API level 27 and 28. In the example, the two versions of the method have the same signature (line 7) and the same method comment (lines 1-6), but different method body, as Android developers added new statements to the method in the API level 28 (lines starting with \texttt{+}). In both API versions, the method returns the type of the SQL statement provided as input.
However, from API level 28, part of the method behavior is changed. In fact,
instead of returning {\small \texttt{STATEMENT\_ABORT}} for any SQL rollback statement as defined in API level 27, the method returns a different value (i.e., {\small \texttt{STATEMENT\_OTHER}}) if the SQL statement aims at rollbacking to a savepoint~\cite{mysql-rollback-2020}.

\vspace{10pt}

\begin{minipage}{0.92\linewidth}
\vspace{5pt}
\begin{lstlisting}[caption={Code snippet extracted by comparing the method \texttt{getSqlStatementType} between Android API level 27 and 28.},label={lst:sample_snippet}]
/**
 * Returns one of the following which represent the type of the given SQL statement.
 * ...
 * @param sql the SQL statement whose type is returned by this method
 * @return one of the values listed above
 */
public static int getSqlStatementType(String sql) {
 String prefixSql = sql.substring(0, 3).toUpperCase(Locale.ROOT);
 else if (prefixSql.equals("ROL")) {
+ boolean isRollbackToSavepoint = sql.toUpperCase(Locale.ROOT).contains(" TO ");
+ if (isRollbackToSavepoint) {
+  ...
+  return STATEMENT_OTHER;
+ }
  return STATEMENT_ABORT;
 }
 return STATEMENT_OTHER;
}}
\end{lstlisting}
\end{minipage}

\vspace{10pt}

Unfortunately, due to this behavioral change, apps that use the method could exhibit compatibility issues. Specifically, an app that relies on the method's behavior as implemented in API level 27 might encounter a failure when running on newer versions of the API. Because API developers did not suitably document the change, app developers might not be aware of the change and hence have a low chance of avoiding such compatibility issue. In the rest of this paper, we present a systematic study that carefully analyzes this family of changes across multiple versions of the Android API and investigates to what extent these changes might affect Android apps.    

\section{Methodology}
\label{sec:methodology}

Our analysis is based on the Android framework codebase. This codebase is one of the largest repository made available on Github and contains over 440,000 commits and nearly one thousand release tags. In this study, we focus our analysis on the ten most recent\footnote{When we started the study in March 2020.} major version releases of the Android framework, as these versions are the ones that are widely used on user devices. (Older releases were less popular and their distribution accounted for less than 3\% of the total distribution.) When selecting a revision for analyzing a major version release, we chose the first release tag associated with each version considered. Table~\ref{tab:droid_history} reports the details of the versions and revisions we considered. Using the versions listed in Table~\ref{tab:droid_history}, we built nine subsequent version pairs (\eg version 19 and 21 constitute a version pair), and use these pairs to identify SEMs.

To the best of our knowledge, no readily-available tool exists to detect SEMs. For this reason, we implemented a prototype tool called \tool{}, which identifies SEMs across different version of the Android framework. At a high-level, the tool takes as inputs the repository containing the codebase of the Android framework and the list of framework version pairs that \tool{} should compare. Given this information, the tool analyzes the version control history of the repository to compare methods across different versions of the Android API. In this step, the tool categorizes a method as a SEM if the method has the same signature, the same method comment, but different method body. The output of \tool{} is a list of SEMs for each version pair analyzed.

\begin{figure}[!t]
	\centering
    \includegraphics[width=0.95\linewidth]{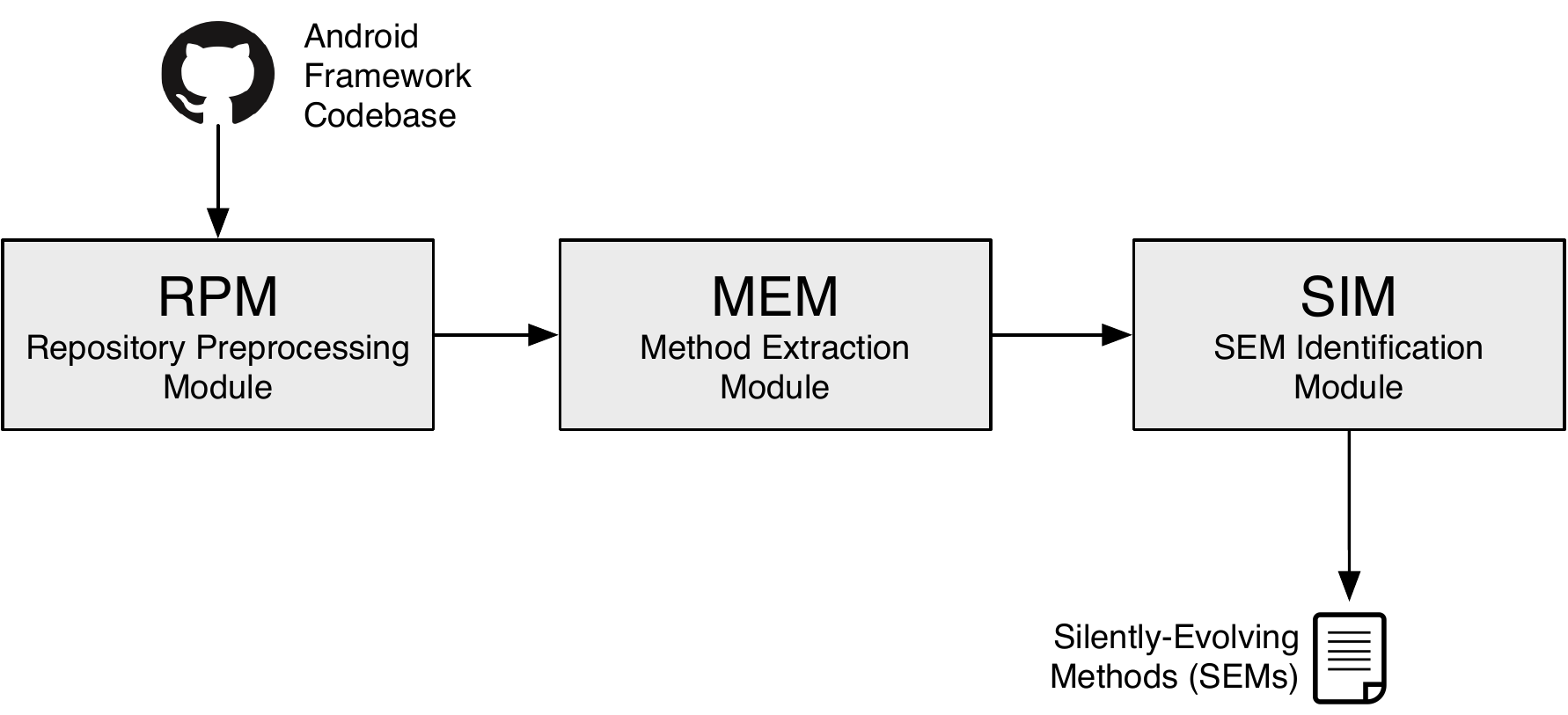}
	\caption{High-level overview of the \tool{} workflow.}
    \label{fig:overview}
    \vspace{-2mm}
\end{figure}

Fig.~\ref{fig:overview} presents a high-level overview of the \tool{} workflow. As the figure highlights, \tool{} uses three modules to identify SEMs. The three modules are the \textit{repository preprocessing module} (RPM), the \textit{method extraction module} (MEM), and the \textit{SEM identification module} (SIM).
We now present the three modules in detail.

\begin{table}[!t]
\centering
\caption{Android Framework revisions used in the study. 
We did not consider API level 20 as this version was focusing on changes for supporting Android Wear.}
\label{tab:droid_history}
\resizebox{0.9\linewidth}{!}{
\begin{tabular}{c l l l}
\hline 
API Level & Code Name   & Release Tag & Distribution \\
\hline
29 & Android10 & android-10.0.0\_r1 & 8.2\% \\
28 & Pie & android-9.0.0\_r1 & 10.4\% \\
27 & Oreo & android-8.1.0\_r1 & 15.4\% \\
26 & Oreo & android-8.0.0\_r1 & 12.9\% \\
25 & Nougat & android-7.1.0\_r1 & 7.8\% \\
24 & Nougat & android-7.0.0\_r1 & 11.4\% \\
23 & Marshmallow & android-6.0.0\_r1 & 16.9\% \\
22 & Lollipop & android-5.1.0\_r1 & 11.5\% \\
21 & Lollipop & android-5.0.0\_r1 & 3.0\% \\
19 & KitKat & android-4.4\_r1 & 6.9\% \\
\hline
\end{tabular}
}
\vspace{-2mm}
\end{table}

\subsection{Repository Prepocessing Module}

Each release of the Android framework codebase contains a large variety of files. In fact, the codebase includes the core implementation of the Android API, the source code of various command-line tools (such as the Android Asset Packaging Tool),
the source code of unit tests, and other types of files.
Because not all of these files are part of the Android API,
the repository prepocessing module analyzes the codebase to identify and select the files that relate to the Android API. The repository preprocessing module performs this task by using a whitelist that we manually constructed, and performs this step for each framework version provided as input to \tool{}. By performing this operation, \tool{} reports only the SEMs that are related to the Android API. \tool{} provides the list of relevant files to the method extraction module for further analysis.

\subsection{Method Extraction Module}
This module locates and extracts relevant information about the Java methods in the source code files of the Android API. Specifically, for each version of the API, the module creates a set of tuples (\textit{apiInfo}) where each tuple (\textit{mInfo}) represents a method in the API and contains the signature (\textit{signature}), the comment (\textit{comment}), and the body (\textit{body}) of the method. The module uses a Java parser to build the abstract syntax tree (AST) for each of the source code files and identifies the API methods in a file by navigating the AST. After locating a method, the module stores the method signature, the method comment, and the method body in \textit{apiInfo}. \tool{} only saves the comments declared through the Javadoc notation (i.e., \texttt{/**...*/}) since only these comments will appear in the documentation of the API. The output of this module are the sets of tuples that are associated with the versions of the API considered.

\subsection{SEM Identification Module}
The SEM identification module identifies SEMs by comparing the relevant methods in the version pairs provided as input to \tool{}.
Algorithm~\ref{algo:comparison} describes how the module identifies SEMs. Given the methods' information from two subsequent versions of the Android API (\textit{apiInfo1} and \textit{apInfo2} in Algorithm~\ref{algo:comparison}), the module iterates over the methods in the versions and compares them (lines~\ref{alg:start}-\ref{alg:stop}). When the algorithm finds methods with matching signatures (\ie they are the same method in different versions of the API), \tool{} first checks whether the methods have a comment associated with them.
If either one of the methods does not have a comment, \tool{} will not consider the methods for further analysis (line~\ref{alg:skip}).
The rationale behind this decision is that we believe that such methods might not be intended for use by app developers as they often resort to the official documentation to learn how to use the API methods.
If both methods have a comment, \tool{} checks whether the comments are the same or not by comparing their text. If the comments have the same text, \tool{} moves forward and compares their method bodies. For simplicity, \tool{} compares the method bodies using the text of their bodies. If the bodies are different, \tool{} categorizes the method as a SEM and adds the method information to the set of SEMs computed for the API versions pair under analysis.




\begin{algorithm}[!t]
\small
\SetAlgoLined
\caption{Detecting Silently-Evolved Methods.}
\label{algo:comparison}
\SetKwInOut{Input}{Input}
\SetKwInOut{Output}{Output}
\Input{\textit{apiInfo1} and \textit{apiInfo2}: method information from two API versions}
\Output{\textit{sems}: set of silently-evolved methods between the two Android versions considered}
 \For{mInfo1 $\in$ apiInfo1} {\label{alg:start}
     \For{mInfo2 $\in$ apiInfo2} {
      \If{mInfo1.signature $\neq$ mInfo2.signature} {
            \textbf{continue}
      }
      \If{mInfo1.comment.isEmpty $||$ mInfo2.comment.isEmpty} {
            \textbf{continue}\label{alg:skip}
      }
      \If{mInfo1.comment $==$ mInfo2.comment}{
        \If{mInfo1.body $\neq$ mInfo2.body}{
            \textit{sems.add(mInfo1, mInfo2)}\label{alg:add}
        }
     }
 }
}\label{alg:stop}
\Return \textit{sems}
\end{algorithm}

\section{Experimental Study}
\label{sec:experiment}

This section discusses our empirical study. In the study, we investigated the following research questions

\begin{itemize}
    \item \textbf{RQ1:} To what extent do SEMs appear in the Android API?
    \item \textbf{RQ2:} What are the characteristics of SEMs?
    \item \textbf{RQ3:} How do SEMs evolve during the development of the Android API?
    \item \textbf{RQ4:} To what extent are PASEMs used in Android apps?
\end{itemize}

In the rest of this section, we answer the research questions by presenting our experimental findings.

\subsection{RQ1: SEMs in the Android API}
With the first research question, we are interested in quantifying the number of SEMs in the Android API.
To this end, we ran \tool{} on the source code of the Android framework codebase using the list of release tags shown in Table~\ref{tab:droid_history}.
\tool{} extracted all the methods in each release and conducted a pairwise comparison between each subsequent version pair (e.g., between \emph{android-4.4\_r1} and \emph{android-5.0.0\_r1}).
Since our study considered ten major version releases of the Android API, \tool{} conducted nine pairwise comparisons.
In total, \tool{} was able to identify 4,769 SEMs and 2,271 of these SEMs are PASEMs.

\begin{table*}[!ht]
    \small
    \centering
    \caption{SEMs identified in our study categorized by their modifiers and annotations.}
    \label{tab:num_apis}
    \begin{adjustbox}{max width=\textwidth}
    \begin{tabular}{c|c|c|c|c|c|c|c|c|c|c|c|c|c|c|c|c|c|c|c|c|c|c|c|c}
    \hline
        \multirow{2}{*}{API Level}   &  \multicolumn{6}{c|}{public}  &  \multicolumn{6}{c|}{default} &  \multicolumn{6}{c|}{protected} &   \multicolumn{6}{c}{private}    \\
        \cline{2-25}
                    &  -  &  static  &  final  &  abstract  &  hide  &  native  &  -  &  static  &  final  &  abstract &  hide   &  native  &  -   &  static  &  final  &  abstract  &  hide  &  native  &  -  &  static  &  final  & abstract  &  hide  &  native \\
       \hline
        19$\to$21    &  502  &  92   &  33  &  1   & 142   &  0  &  38  &  6  &  1  &  0  &  6  &  0  &  16  &  0  &  1  &  0  & 24  &  0  &  52  &  5  &  3  &  0  &  39  &  0 \\ 
        21$\to$22    &  112  &  18   &  4   &  125 &  80   &  0  &  16  &  2  &  1  &  0  &  1  &  0  &  1   &  0  &  1  &  4  &  7   &  0  &  28  &  4  &  1  &  0  &  14  &  0 \\
        22$\to$23    &  279  &  75   &  49  &  0   &  171  &  9  &  15  &  9  &  1  &  0  &  4  &  0  &  7   &  0  &  0  &  0  &  4   &  0  &  63  &  10 &  1  &  0  &  28  &  0 \\
        23$\to$24    &  436  &  80   &  28  &  0   &  336  &  0  &  16  &  9  &  2  &  0  &  5  &  0  &  5   &  0  &  0  &  0  &  4   &  0  &  44  &  13 &  0  &  0  &  15  &  0 \\
        24$\to$25    &  26   &  9    &  5   &  0   &  37   &  0  &  4   &  0  &  0  &  0  &  1  &  0  &  0   &  0  &  0  &  0  &  3   &  0  &  3   &  3  &  0  &  0  &  2  &  0 \\
        25$\to$26    &  233  &  60   &  14  &  0   &  200  &  0  &  19  &  4  &  5  &  0  &  24  &  0  &  3   &  0  &  0  &  0  &  7   &  0  &  32  &  9  &  2  &  0  &  13  &  0 \\
        26$\to$27    &  64  &  25   &  3   &  0   &  193  &  0  &  5   &  2  &  0  &  0  &  5  &  0  &  0   &  0  &  0  &  0  &  0   &  0  &  7   &  2  &  0  &  0  &  2  &  0 \\
        27$\to$28    &  145  &  111  &  15  &  0   &  151   &  0  &  9  &  3  &  0  &  0  &  4  &  0  &  0   &  0  &  0  &  0  &  4   &  0  &  19  &  1  &  0  &  0  &  3  &  0 \\
        28$\to$29    &  166  &  63   &  11  &  0   &  263  &  0  &  4  &  1  &  0  &  0  &  14  &  0  &  2   &  0  &  0  &  0  &  6   &  0  &  29  &  4  &  0  &  0  &  6  &  0 \\
        \hline
    \end{tabular}
    \end{adjustbox}
\end{table*}

\begin{figure*}[!ht]
    \centering
    \subfigure[]{
        \centering
        \includegraphics[width=0.31\textwidth]{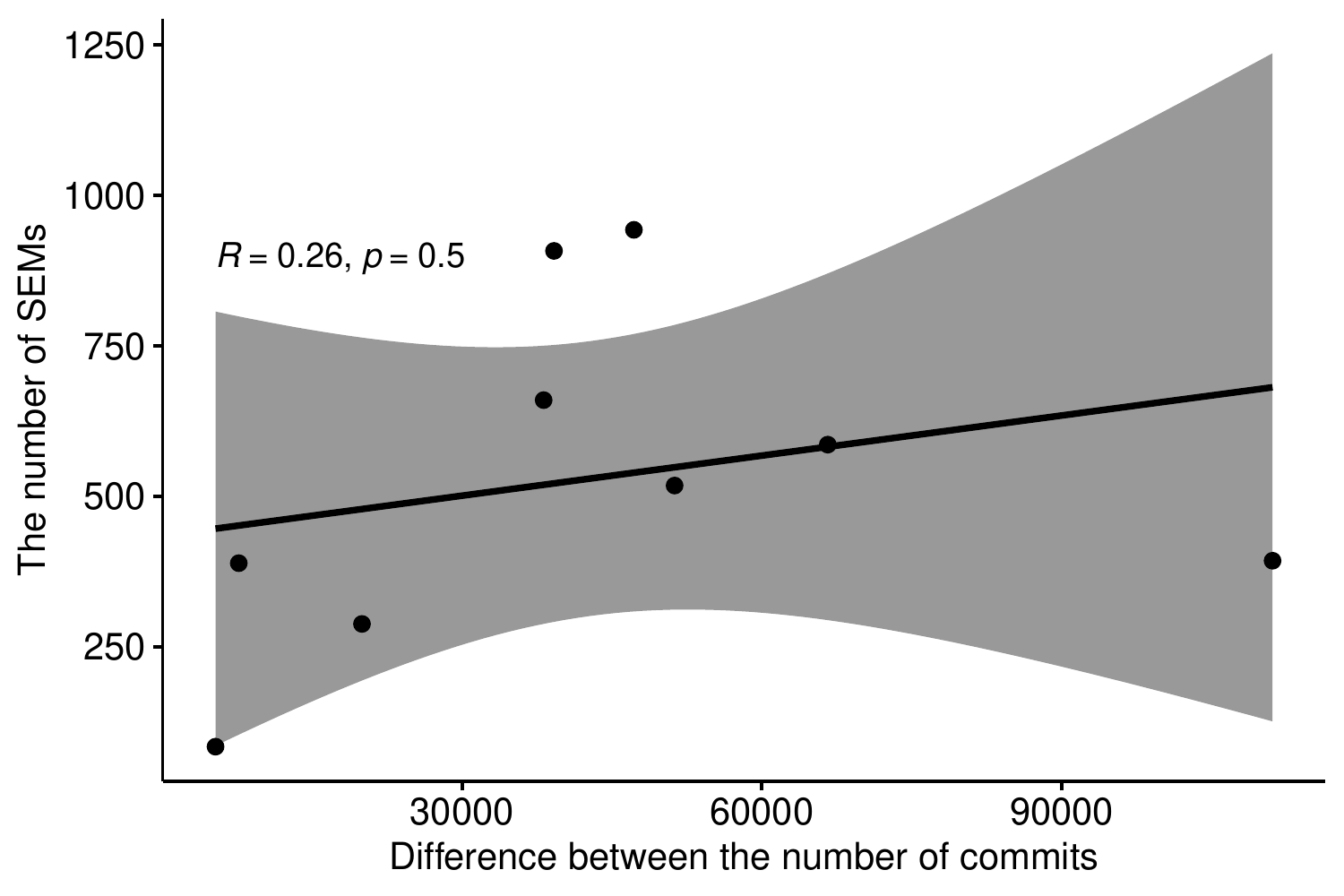}
        \label{fig:sem_commit}
    }
    \hfill
    \subfigure[]{
        \centering
        \includegraphics[width=0.31\textwidth]{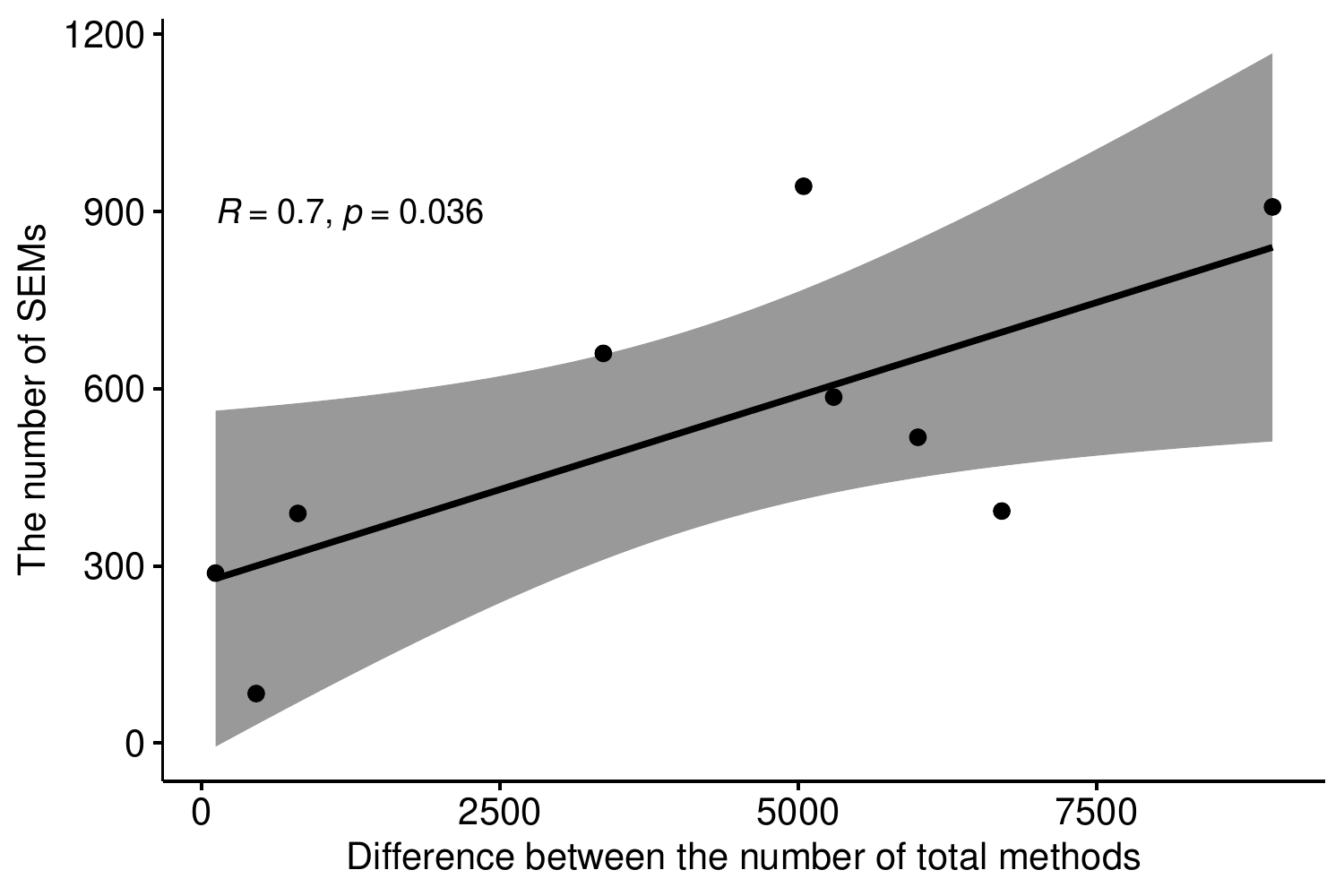}
        \label{fig:sem_total_method}
    }
    \hfill
    \subfigure[]{
        \centering
        \includegraphics[width=0.31\textwidth]{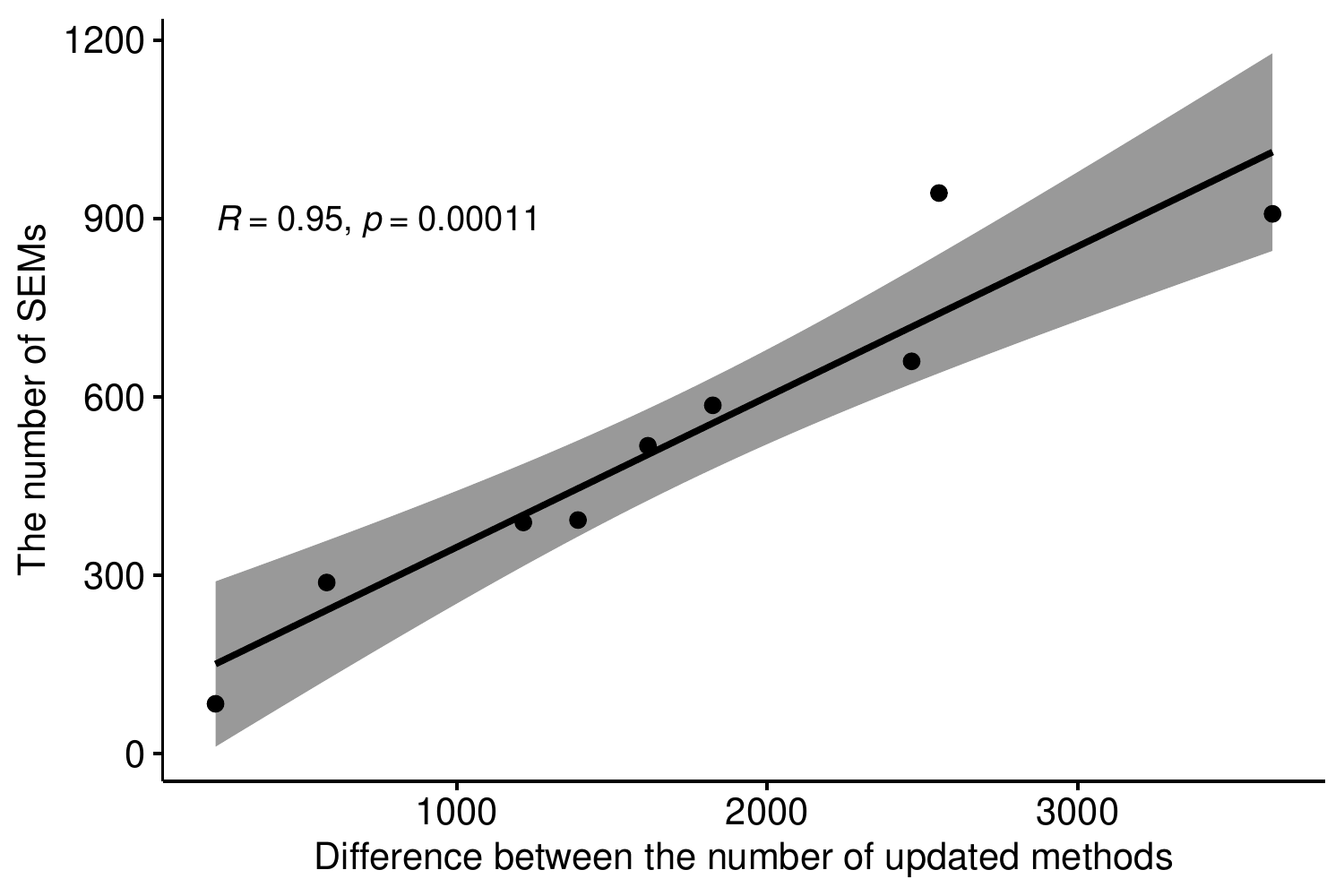}
        \label{fig:sem_upd_method}
    }
    \caption{Correlations between the number of SEMs and the differences of the number of commits, total methods, and updated methods in two subsequent releases, respectively.}
    \label{fig:corr_commit_methods}
\end{figure*}

After identifying SEMs, we also analyzed the modifiers and annotations associated with the methods to determine the potential impact of the methods on client apps. Table~\ref{tab:num_apis} reports the number of SEMs identified by \tool{} and categorizes them by their modifiers and annotations. The table is divided into five sections. The first section (API Level) reports the information of the version pair considered in the study. The other four sections group methods according to their Java access modifier. In Table~\ref{tab:num_apis}, the columns labeled with the symbol `-' report the number of SEMs that have the Java access modifier as their only modifier. Furthermore, if a method has the Java access modifier and multiple additional modifiers or annotations, we counted the method in all the columns that apply. For example the SEM {\small \texttt{getInstance}} from the {\small \texttt{ConnectivityManager}} class in API level 23 is a {\small \texttt{private}} method that has both the {\small \texttt{static}} modifier and the {\small \texttt{hide}} annotation. In this case, we counted the method both in the ``static'' and the ``hide'' columns of the ``private'' section.

The majority of SEMs are declared as public (i.e., they are PASEMs) and these methods can be directly accessed by of Android apps.
Based on the update types of PASEMs, these apps may be subject to compatibility issues, which can lead to field failures if the update has changed the method's semantics (e.g., the API method presented in our motivating example and reported in Listing~\ref{lst:sample_snippet}).
We believe that API developers should pay particular attention to updating these method comments so that app developers can suitably account for the semantic changes affecting their apps.

Although the number of version pairs considered is on the low side (and this might affect the validity of the results), we performed a correlation analysis on the SEMs we identified. Specifically, Fig.~\ref{fig:corr_commit_methods} presents the correlation (obtained via Pearson's correlation coefficient) between the total number of SEMs and the difference between the number of commits (Fig.~\ref{fig:corr_commit_methods} (a)), the number of methods including updated methods (Fig.~\ref{fig:corr_commit_methods} (b)), and the number of updated methods (Fig.~\ref{fig:corr_commit_methods} (c)) in each version pair considered.
These correlation results (i.e., Pearson's correlation coefficient $R$ and $p-value$) show that the introduction of SEMs is not strongly correlated with the number of Github commits (which do not necessarily lead to method changes) but strongly correlated with the difference in the number of methods, especially the difference in the number of updated methods, in two subsequent releases.

\begin{tcolorbox}[title=\textbf{Answer to RQ1}, left=2pt, right=2pt,top=2pt,bottom=2pt]
Based on our empirical results, we can confirm that SEMs are present in the Android API. Considering ten major version releases of the API, we were able to identify 4,769 SEMs, including 2,271 PASEMs. These PASEMs could lead to compatibility issues in their client Android apps. Furthermore, the more methods are updated in a version release, the more SEMs can be introduced in the API.
\end{tcolorbox}

\subsection{RQ2: Understanding SEMs}

In the second research question, we are interested in understanding the main purposes behind the updates of SEMs.
Specifically, we would like to check whether the updates are related to simple code refactorings that would introduce no harm to the system or involved in semantic changes that could break the execution of existing Android apps.
Ideally, we would expect that SEMs should only involve code refactorings such as renaming variables and attributes.
They should not include semantic changes as those changes could break the execution of client apps.
The comments, especially the Javadocs, of the corresponding methods should have subsequently been updated to properly advise app developers to update their apps so as to be aligned with the changed APIs' new semantics.

\begin{table*}[!ht]
    \centering
    \caption{Manual classification on selected sample PASEMs.}
    \label{tab:manual_class}
    \begin{adjustbox}{max width=\textwidth}
        \begin{tabular}{c|l|l|l|l|l|l|l|cl}
        \hline
                \multirow{2}{*}{API Level}    &  \multirow{2}{*}{Sample}  &  \multicolumn{4}{c|}{Update type} & \multirow{2}{*}{Refactoring}  &  \multirow{2}{*}{Uncertain}   &    \multirow{2}{*}{Disagreement}   \\
                \cline{3-6}
                & & Added  &  Removed & Changed & Sum & & & & \\
        \hline
              19$\to$21  &   83   &  16  &  2  & 34   &   52 (62.65\%)   &  22 (26.51\%)  &  9 ~(10.84\%)  &  22 (26.51\%)  \\
              21$\to$22  &   54   &  17  &  1  & 14   &   32 (59.26\%)   &  19 (35.19\%)  &  3 ~(5.56\%)   &  1 ~(1.85\%)   \\
              22$\to$23  &   76   &  23  &  1  & 27   &   51 (67.10\%)   &  21 (27.63\%)  &  4 ~(5.26\%)   &  21 (27.63\%)  \\
              23$\to$24  &   81   &  29  &  1  & 35   &   65 (80.25\%)   &  13 (16.05\%)  &  3 ~(3.70\%)   &  9 ~(11.11\%)  \\
              24$\to$25  &   25   &  8   &  1  & 11   &   20 (80.00\%)   &  2 ~(8.00\%)   &  3 ~(12.00\%)  &  5 ~(20.00\%)  \\
              25$\to$26  &   71   &  15  &  2  & 24   &   41 (57.75\%)   &  23 (32.39\%)  &  7 ~(9.83\%)   &  11 (15.49\%)  \\
              26$\to$27  &   42   &  13  &  0  & 15   &   28 (66.67\%)   &  14 (33.33\%)  &  0 ~(0.00\%)   &  2 ~(4.76\%)   \\
              27$\to$28  &   65   &  9   &  0  & 21   &   30 (46.15\%)   &  24 (36.92\%)  &  11 (16.92\%)  &  12 (18.46\%)  \\
              28$\to$29  &   65   &  13  &  2  & 29   &   44 (67.69\%)   &  10 (15.38\%)  &  11 (16.92\%)  &  14 (21.53\%)  \\
        \hline
              Total      &   562  &  143 &  25 & 199  &   363 (64.59\%)  &  148 (26.33\%) &  51 ~(9.07\%)  &  97 (17.26\%)  \\
        \hline
        \end{tabular}
    \end{adjustbox}
\end{table*}

\begin{lstlisting}[caption={An example of PASEM flagged as code refactoring.},label={lst:refactoring}, language=Java, frame=shadowbox, floatplacement=H]
// Between android-8.1.0_r1 and android-9.0.0_r1
public CharSequence[] getTextArray(@StyleableRes int index) {
  final TypedValue value = mValue;
- if (getValueAt(index * AssetManager.STYLE_NUM_ENTRIES, value)) {
+ if (getValueAt(index * STYLE_NUM_ENTRIES, value)) {
\end{lstlisting}

Unfortunately, to the best of our knowledge, our community has not yet made available tools for effectively determining whether an update of method code (i.e., method diff) is related to semantic change or not. 
To this end, in this work, we resort to manual efforts to classify the purposes behind the updates of PASEMs (i.e., the nature of the changes).
We choose to classify PASEMs instead of SEMs because only PASEMs could directly impact the execution of client apps available in the wild.
Since manual efforts are known to be time-intensive, it becomes impractical for us to manually classify all the PASEMs identified previously. To that end, we resort to randomly sample a set of PASEMs to fulfill the purpose.
To ensure that the sampled PASEMs are representative, we turn to the well-known online Sample Size Calculator\footnote{https://www.surveysystem.com/sscalc.htm} to determine the number of PASEMs for manual classification (with a confidence level of 95\% and margin of error of 10\%).

The second column in Table~\ref{tab:manual_class} enumerates the number of samples randomly selected from each framework iteration.
For each method, two of the authors of the paper independently categorized the method as either \emph{refactoring} or \emph{semantic change}.
If the authors cannot make a decision in 10 minutes, the corresponding PASEM will be flagged as uncertain.
After completing the independent manual classification, the two authors then set up meetings to discuss their decisions until consensuses reached.
The final results are summarized in the last four columns in Table~\ref{tab:manual_class}. 
In the manual classification, only 97 out of 562 are classified as different types by the two different authors achieving a high inter-rater reliability of 82.74\%. After the meeting, 41 out of 97 are concluded as semantic changes accounting for 42.27\% while the refactorings and uncertains are made up of 23.71\% and 34.02\% respectively.
Surprisingly, all in all, slightly more than a quarter of the randomly selected PASEMs are related to code refactorings (Listing~\ref{lst:refactoring} presents such an example), and around two-thirds of the randomly selected PASEMs are related to semantic changes (e.g., logic added, removed, or changed that is complicated update including statements added and removed).
This result shows that SEMs are a severe problem in the Android framework.
The framework maintainers should pay special attention to carefully handle these methods, i.e., avoid introducing SEMs in future releases and document the semantically changed methods.


\begin{lstlisting}[caption={An example of PASEM flagged as uncertain.},label={lst:uncertain}, floatplacement=H]
// Between android-9.0.0_r1 and android-10.0.0_r1
public void setVolumeTo(int value, int flags) {
  try {
- mSessionBinder.setVolumeTo( mContext.getPackageName(), mCbStub, value, flags);
+ // Note: Need both package name and OP package name. Package name is used for
+ // RemoteUserInfo, and OP package name is used for AudioService's internal
+ // AppOpsManager usages.
+ mSessionBinder.setVolumeTo( mContext.getPackageName(), mContext.getOpPackageName(), mCbStub, value, flags);
  } catch (RemoteException e) {
\end{lstlisting}

As shown in Table~\ref{tab:manual_class} (the fifth column), around 10\% of PASEMs are flagged as uncertain.
The majority of those methods are related to updates of callee methods, which may further involve complicated changes.
Listing~\ref{lst:uncertain} presents such an example.
The original callee method \emph{setVolumeTo()} called by \emph{mSessionBinder} has been replaced by a new one that involves a new parameter, which is not needed by the original version.
The callee method is only defined in a Java interface called \emph{ISessionController}, which is non-trivial for the authors to manually identify its dynamically bound object (in a short time), considering that the Android framework is one of the most complicated open-source projects.
Therefore, this update is flagged as \emph{uncertain}.

\begin{tcolorbox}[title=\textbf{Answer to RQ2}, left=2pt, right=2pt,top=2pt,bottom=2pt]
Our manual classification reveals that the majority of PASEMs (over 64.59\%) do involve semantic changes, and such changes may involve complicated updates of the code.
\end{tcolorbox}

\subsection{RQ3: Evolution of PASEMs}

\begin{figure}[!t]
	\centering
    \subfigure[SEMs]{
        \centering
        \setlength{\abovecaptionskip}{0.cm}
        \includegraphics[width=0.36\linewidth]{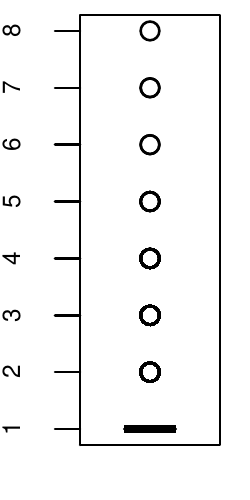}
        \label{fig:sem_upds}
    }%
    \hspace{16pt}
    \subfigure[PASEMs]{
        \centering
        \setlength{\abovecaptionskip}{0.cm}
        \includegraphics[width=0.36\linewidth]{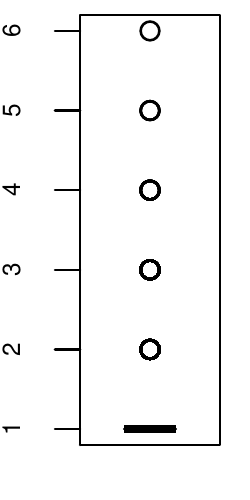}
        \label{fig:sea_upds}
    }
	\caption{Distribution of update times of SEMs from level 19 to 29.}
    \label{fig:method_upd_times}
\end{figure}

In this research question, we are interested in exploring the evolution of SEMs under the evolution of the Android framework codebase.
To this end, we first look at the number of times a given method is silently changed.
Fig.~\ref{fig:method_upd_times} illustrates the distribution of such times for all the identified 4,769 SEMs and 2,271 PASEMs.
Expectedly, the majority of methods (61.12\%) that are flagged as SEMs are only silently evolved once, in the meanwhile, 79.03\% of PASEMs are silently evolved once, suggesting that SEMs might not be intentionally introduced by the framework developers.
Nevertheless, there are several methods that have indeed been repeatedly changed silently. 
For example, method \emph{loop} of file \texttt{core/java/android/os/Looper.java} has been silently updated six times, among the nine considered iterations. The class Looper is used to run a message loop for a thread. For the specific method, it is actually a static public method that is used to run the message queue in this thread. 

Among the 4,769 SEMs, interestingly, only 80 of them have their comments updated along with the update of the implementation in the following up revisions of the Android framework.
This experimental result suggests that framework maintainers might not yet be made aware of SEMs or at least are not well-motivated to mitigate the introduction of SEMs.  
Listing~\ref{lst:upd_comment} illustrates one of such updated comments.
The implementation of the method has been updated in the release android-5.0.0\_r1 while its comments are only updated at the revision android-6.0.0\_r1.
Nevertheless, although rare, the fact that some of the SEMs are indeed resolved by the framework maintainers shows that SEMs are indeed a problem that should be carefully addressed. 

\begin{minipage}{0.92\linewidth}
\medskip
\begin{lstlisting}[caption={The comment of SEM \emph{setCheckMarkDrawable} is updated in a future release while its body is not changed.}, label={lst:upd_comment}, language=Java, frame=shadowbox, floatplacement=H]
//android.widget.CheckedTextView.setCheckMarkDrawable
 /**
- * Set the checkmark to a given Drawable. This will be drawn when {@link #isChecked()} is true.
- * @param d The Drawable to use for the checkmark.
+ * Set the check mark to the specified drawable.
+ * <p>
+ * When this view is checked, the drawable's state set will include
+ * {@link android.R.attr#state_checked}.
+ * @param d the drawable to use for the check mark
+ * @attr ref android.R.styleable#CheckedTextView_checkMark
  * @see #setCheckMarkDrawable(int)
  * @see #getCheckMarkDrawable()
- * @attr ref android.R.styleable#CheckedTextView_checkMark
  */
\end{lstlisting}
\end{minipage}



\begin{figure}[t!]
	\centering
    \includegraphics[width=0.8\linewidth]{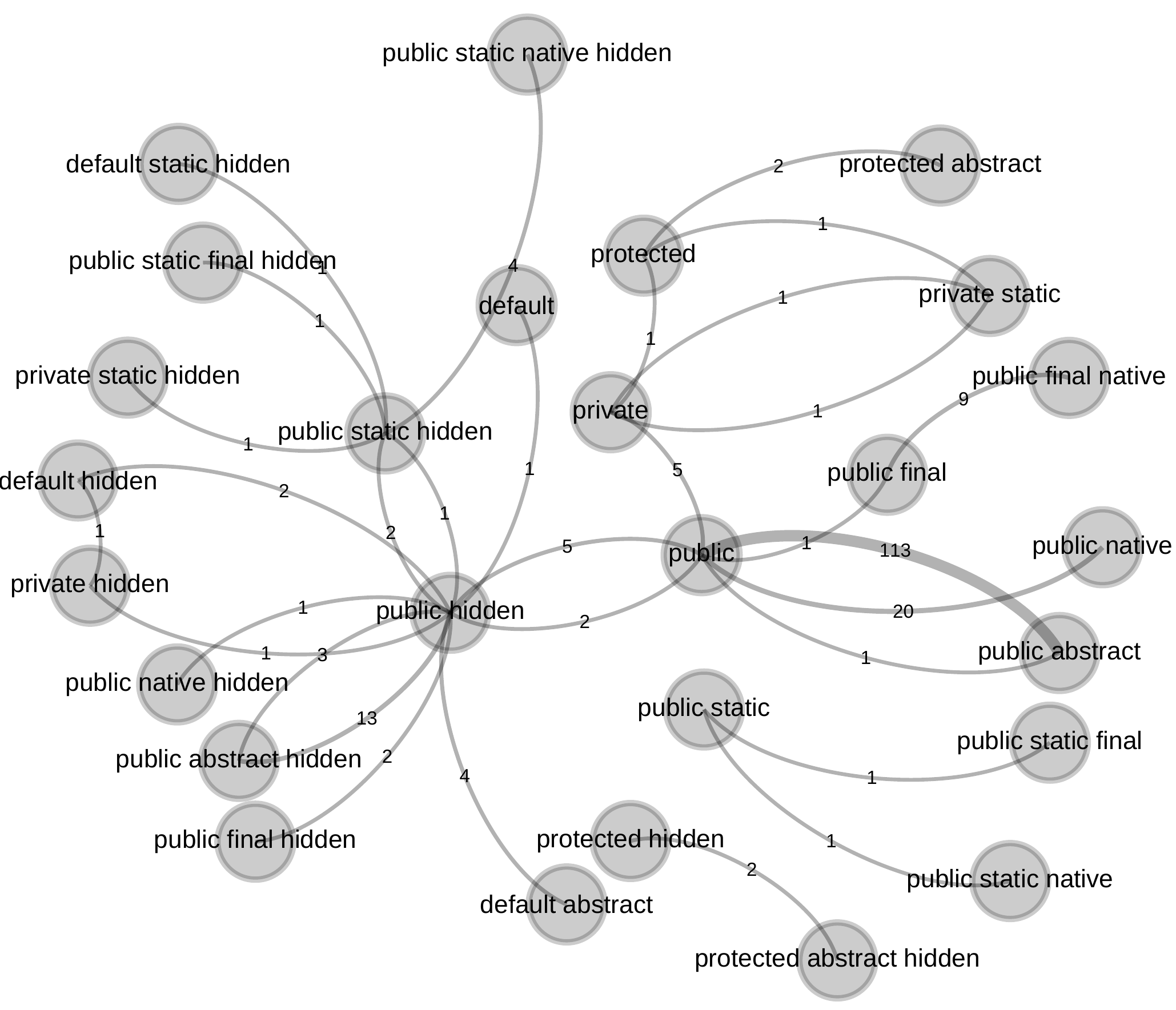}
	\caption{The update of modifiers. The line in clockwise shows the direction of the update and the weight on the line shows the number of SEMs having such updates.}
    \label{fig:modifier_update}
    \vspace{-5mm}
\end{figure}

We further look at the evolution of method modifiers for all the identified SEMs.
Fig.~\ref{fig:modifier_update} summarizes such results in a directed graph.
Each node represents a distinct modifier type, while each arrowed edge represents an evolution, e.g., from a \emph{src} modifier to the \emph{dest} modifier.
The weight of each edge subsequently represents the number of times the corresponding evolution happens during the evolution of the Android framework codebase.

In total, we have observed 204 times of modifier changes.
Expectedly, none of the changes are made from \emph{public} to other low-level accessibilities (except one from \emph{public static hidden} to \emph{private static hidden}\footnote{Hidden APIs cannot be directly accessed by Android apps.}) as such changes would break the compilation rules for Android apps, resulting in explicit compatibility issues.
Interestingly, the majority of modifier changes are related to altering \emph{non-abstract} methods to \emph{abstract} methods or exchanging \emph{native} methods with \emph{non-native} methods.
Listing~\ref{lst:mod_upd} represents an example of modifiers update from \emph{public} to \emph{public abstract} extracted between android-5.0.0\_r1 and android-5.1.0\_r1. The class, the method belongs to, is actually an abstract class.
The body of the method is to provide the default behavior, which is to throw \texttt{MustOverrideException} if the subclass extends the \emph{WebSettings} but does not explicitly override the method. The update of the method reinforces the need to override the method by declaring it as \emph{abstract}, which utilizes the compile-time check to ensure that the method will be overridden in the subclass.
Although this type of change is trivial, it will theoretically cause compatibility issues for existing Android apps.
Indeed, previously, even if developers do not override the API method \emph{setBuiltInZoomControls()} when extending the \emph{WebSettings} class, grammatically speaking, there will be no compile error.
However, with the latest SDK version, the accessibility of \emph{setBuiltInZoomControls()} has changed to \emph{public abstract}, there will be a compile error if the method is not explicitly overridden.
Therefore, ideally, this type of change should be avoided by framework maintainers, and subsequently should be considered by software analyzers aiming at detecting compatibility issues in Android apps.


\begin{minipage}{0.92\linewidth}
\medskip
\begin{lstlisting}[caption={An example of modifier update from public to public abstract.},label={lst:mod_upd}, language=Java, frame=shadowbox, floatplacement=H]
// This is an abstract base class: concrete WebViewProviders must
// create a class derived from this, and return an instance of it in the
// WebViewProvider.getWebSettingsProvider() method implementation.
public abstract class WebSettings {
- public void setBuiltInZoomControls(boolean enabled) {
-  throw new MustOverrideException();
- }
+ public abstract void setBuiltInZoomControls(boolean enabled);
\end{lstlisting}
\end{minipage}

The majority of the updates related to native methods share the same change pattern, which is to provide a wrapper method with the same name while calling the native method to provide the same behavior (cf.  (Listing~\ref{lst:mod_native}).
Even though the modifiers were updated from \emph{public native} to \emph{public}, the signatures of the APIs provided to developers are not changed. Practitioners hence can still use the same method signature to implement their intentions. Therefore, this type of update will unlikely to introduce compatibility issues into running Android apps.

\begin{minipage}{0.92\linewidth}
\medskip
\begin{lstlisting}[caption={An example of modifier update from public native to public. },label={lst:mod_native}, floatplacement=H]
// Code snippet from android-4.4_r1
94  public native int getHeight();
// Code snippet from android-5.0.0_r1
107 public int getHeight() {
108  return nativeGetHeight(mNativePicture);
109 }
\end{lstlisting}
\end{minipage}


\begin{tcolorbox}[title=\textbf{Answer to RQ3}, left=2pt, right=2pt,top=2pt,bottom=2pt]
The majority of SEMs is only introduced into the framework once, without following up updates. Moreover, SEMs may involve updating the method's modifiers that could further introduce runtime issues to client Android apps.
\end{tcolorbox}

\subsection{RQ4: PASEMs in Android apps}

Since PASEMs may involve semantic changes, their client apps may suffer from incompatible issues (because of those silent semantic changes) when they are running on devices with different SDK versions.
Therefore, in this last research question, we are interested in knowing if PASEMs are used by Android apps. If so, to what extent are they accessed, and what are the potential impacts such usages could bring to Android app developers?
To answer these questions, we introduce a simple app scanner to the community, which takes as input an Android APK and the list of PASEMs identified in this work and outputs the list of PASEMs that are actually accessed by the APK.
The app scanner is implemented on top of the famous Soot analysis framework~\cite{lam2011soot}.
The APIs are identified at the Jimple level, which is one of the intermediate representation types provided by Soot to ease the analysis of Android apps.

\begin{figure}[!t]
	\centering
	\subfigure[Total PASEMs.]{
	    \includegraphics[width=0.4\linewidth]{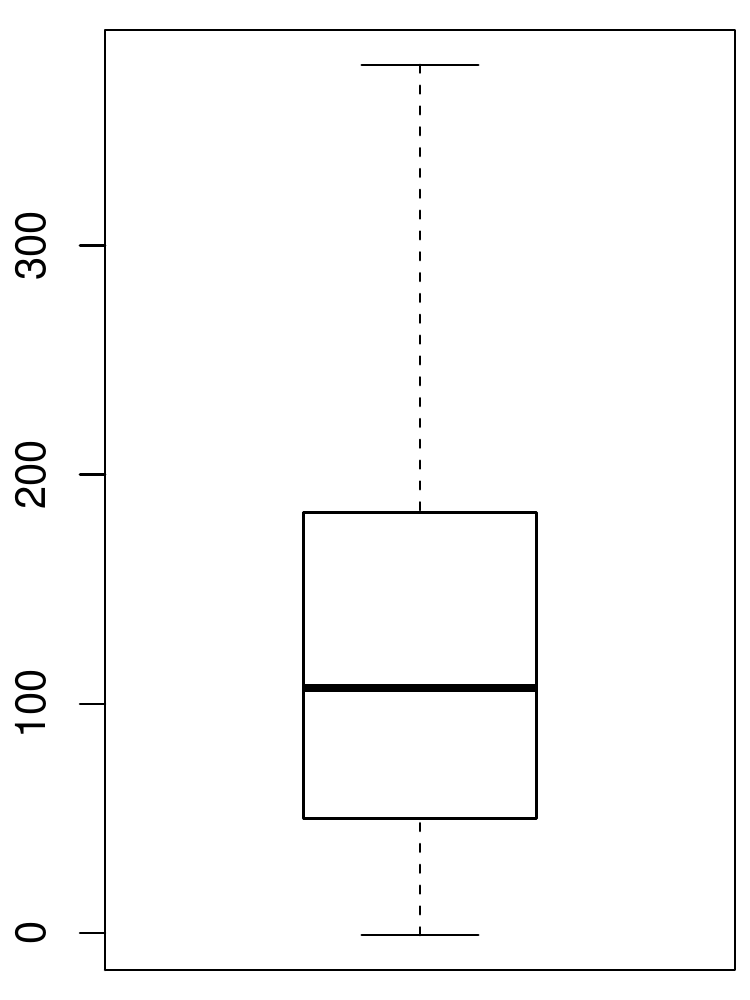}
	    \label{fig:seas_stats_2020}
	}%
	\hspace{20pt}
	\subfigure[Semantically updated PASEMs.]{
	    \includegraphics[width=0.4\linewidth]{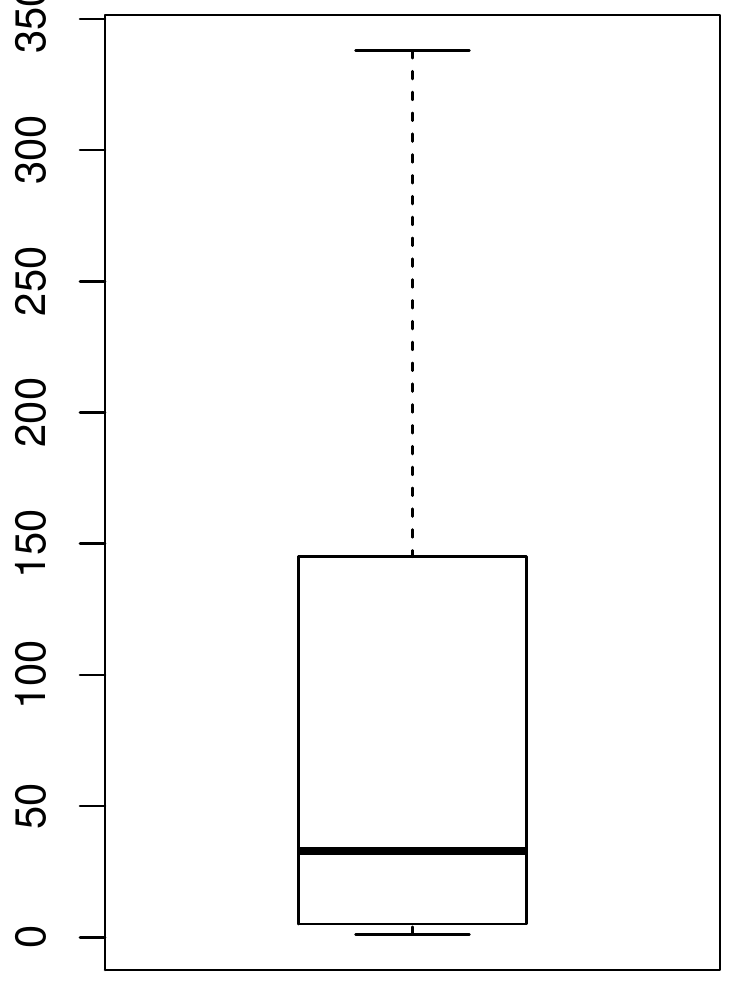}
	    \label{fig:seas_sample_2020}
	}
	\caption{The distribution of the number of PASEMs used in a set of randomly selected 1,000 APKs published on 2020.}
    \label{fig:sea_apk_2020}
    \vspace{-1mm}
\end{figure}

Specifically, to fulfill our experiments, we randomly select 1,000 apps published in 2020 from the official Google Play store.
All of the selected apps are then sent to the aforementioned app scanner to check whether they have leveraged PASEMs or not.
Interestingly and surprisingly, among the 1,000 apps, 957 of them have accessed PASEMs.
Moreover, in total, 44.25\% PASEMs are accessed.
Fig.~\ref{fig:seas_stats_2020} further presents the distribution of the number of PASEMs accessed by these apps, giving a median and average number of PASEMs at 107 and 126, respectively.
This experimental result shows that PASEMs have been significantly accessed by Android apps, which could subsequently suffer from ``hidden'' incompatible issues.
This result strongly suggests that the Android framework maintainers should pay special attention to avoid introducing PASEMs.
This finding is further backed up by the fact that semantically updated PASEMs are also significantly accessed by Android apps, as illustrated in Fig.~\ref{fig:seas_sample_2020}, for which only the 363 manually confirmed PASEMs (involving semantic changes) are considered.

Previous studies, such as the one proposed by He et al.~\cite{he2018understanding}, shows that the convention to address the evolution-induced API compatibility issue is to add additional conditions to check the practical running API level of the device (e.g., through the default constant value of \texttt{VERSION.SDK\_INT}) before invoking the corresponding APIs. 
In this work, we try to detect how many different PASEMs are called after the API level condition check, where the PASEMs are referred to as protected PASEMs. 
Interestingly, 469 out of 1,000 Apps indeed contain protected PASEMs, for which there are 134 PASEMs called under protection. 
Table~\ref{tab:top10_1k} presents the top-10 PASEMs ranked by their protection times.
The fact that the top-ranked PASEMs are protected more than 800 times shows that PASEMs could be protectively accessed many times within the same app.


Moreover, we go one step deeper to check the actual API levels leveraged for protecting the invocation of PASEMs.
Our experimental results reveal that a large portion of PASEMs (i.e., 45 or 33.58\%) are protected by different API levels, indicating possible errors of app developers, although they have attempted to protect the accessed PASEMs.
The possible reasons behind these disagreements between developers could be that PASEMs are silently (hence hiddenly) introduced to the framework.
There is hence no documentation for developers to correctly use these PASEMs.
Subsequently, developers have to independently identify the API levels suitable for protecting PASEMs independently, based on their empirical evidence.
The fact that developers have attempted to protect the access of PASEMs (although may incorrectly do so because of lacking documentation) indicates that PASEMs can indeed introduce runtime issues to Android apps.
Therefore, we argue that PASEMs should be totally avoided by framework maintainers.
There is also a need to introduce automated tools to regulate that.
Our approach \tool{} could be leveraged to achieve such a purpose.


\begin{table}[!t]
    \centering
    \scriptsize
    \caption{The top ten protected PASEMs in 1,000 APKs}
    \begin{adjustbox}{max width=\linewidth}
    \begin{tabular}{l|l}
        \hline
        PASEM  &  times \\
        \hline
        android.content.res.TypedArray: void recycle() & 848 \\
        android.os.Bundle: void putParcelable(String,Parcelable) & 788 \\
        android.os.Bundle: void putCharSequence(String,CharSequence) & 623 \\
        android.text.TextUtils: void writeToParcel(CharSequence,Parcel,int) & 384 \\
        android.app.AlarmManager: void setExact(int,long,PendingIntent) & 359 \\
        android.app.AlarmManager: void setExactAndAllowWhileIdle(int,long,PendingIntent) & 289 \\
        android.app.Activity: boolean navigateUpTo(Intent) & 286 \\
        android.app.Activity: void startIntentSenderForResult(IntentSender,int,Intent,int,int,int,Bundle) & 285 \\
        android.view.ViewGroup: void removeView(View) & 283 \\
        android.os.Bundle: void putAll(Bundle) & 273 \\
        \hline
    \end{tabular}
    \end{adjustbox}
    \label{tab:top10_1k}
    \vspace{-3mm}
\end{table}





\begin{tcolorbox}[title=\textbf{Answer to RQ4}, left=2pt, right=2pt,top=2pt,bottom=2pt]
PASEMs have indeed commonly been accessed by real-world Android apps. 
Some of them are even accessed with protections (by checking the running API level), indicating that practitioners have realized that those PASEMs could introduce runtime issues to their apps, although they are not documenting.
\end{tcolorbox}

\section{Discussion}
\label{sec:discussion}
We now discuss the potential implications of this study for both practitioners and our fellow researchers, as well as some promising future research directions that could be built on the findings of our research (cf.~\ref{subsec:implication}). 
After that, we present the potential threats to the validity of this study (cf. Section~\ref{discussion_threat}).

\subsection{Implication for Practitioners and Researchers}
\label{subsec:implication}

As shown in Figure~\ref{fig:sea_apk_2020}, the usages of PASEMs are common in real-world Android Apps, and over 60\% of the PASEMs involve truly semantic changes, as demonstrated in Table~\ref{tab:manual_class}. As semantic changes could introduce potential crashes (or security, efficiency issues) in daily use of the Android Apps, we argue that framework maintainers should try their best to avoid introducing PASEMs. This should also apply to the maintenance of any other third-party frameworks or libraries that provide APIs to facilitate the development of client apps.
Subsequently, the client app developers should pay special attention to those silently evolved methods when developing their apps.

As shown in Table~\ref{tab:manual_class}, slightly more than a quarter of SEMs do not involve semantic changes but are simply related to code refactorings.
When performing compatibility analyses, these methods could be ignored as they will not introduce runtime issues to their users (i.e., client apps).
However, to the best of our knowledge, our community has not introduced promising tools to automatically decide if a given code diff involves semantic changes. State-of-the-art refactoring detection tools~\cite{tsantalis2020refactoringminer, falleri2014fine} are not capable of accurately achieving that. Therefore, we argue that there is a strong need to invent an automated approach to locate semantic changes during the evolution of software systems.
With the help of this tool, SEMs with semantic changes could be automatically identified and thereby mitigated by codebase maintainers.

Besides the implementation update of the methods, the methods' modifiers might also be updated along with. The update of the modifiers also could bring in big problems, such as the example in listing~\ref{lst:mod_upd}.
Therefore, we argue that our fellow practitioners should pay attention to the updates of methods' modifiers when updating their software systems.

\subsection{Threats to Validity}
\label{discussion_threat}

As it is the case of most empirical evaluations, there are threats to validity associated with the results we presented. In terms of external validity, our results might not generalize to other version pairs in the Android framework. In particular, we identified SEMs in only nine versions pairs. This limitation is an artifact of the complexity associated with the manual analysis performed in the study. To mitigate this threat, we considered all the widely-used major version releases of the Android framework. In relation to external threats to validity, the results presented for answering RQ4 might not generalize to other apps. We mitigated this threat by randomly selecting apps from AndroZoo, one of the most comprehensive app datasets made available to the research community. Finally, an additional threat to validity could be posed by the fact that our study involves manual tasks, which could have introduced errors in the results we presented. For example, in our study, we had to manually inspect and summarize the code associated with SEMs. To mitigate this threat, two authors cross-validated and inspected the results.



\section{Related Work}
\label{sec:related}


\subsection{API evolution}
While API evolves to meet new feature requirements, to fix bugs etc., developers need to update their implementation of the Application and publish the newer version to provide a stable running environment for their customers. McDonnell et al.~\cite{mcdonnell2013empirical} conducted an extensive empirical study on the stability and adoption of Android APIs while focusing on the relationship between the API evolution and client adoption. The authors in the paper confirm that the Android API evolves more frequently than the client adoption and what's more, the more frequent update of the APIs the longer time for client to adopt the ones. Bavota et al.~\cite{bavota2014impact} and Linares-V{\'a}squez~\cite{linares2013api} studied the relationship between the popularity of the Android Applications and the stability of the SDK APIs. Their empirical study reveal that the more enjoyable Android Apps are prone to call the less updated APIs. Works~\cite{li2018cid} and~\cite{he2018understanding} investigate the API-related compatibility issues. Li et al.~\cite{li2018cid} present an approach CiD to highlight the API usage that can lead to potential compatibility issues by analysing the framework release history and identifying the methods without API level checking. He et al.~\cite{he2018understanding} investigate the evolution-induced compatibility issues in Android Applications. Their research shows that the Android Support library only provides limited support for the new APIs in each release and the majority of the Applications need to handle the evolution-induced compatibility issues in their own implementation. Different from the existing work focusing on the general APIs, what we do is to disclose the silently evolved methods that always ignored by developers and researchers.

\subsection{API pattern}
In addition, researchers propose many different approaches to detect Android malware to address the security problems of remote control, privilege escalation, and privacy leakage etc. Chan et al.~\cite{chan2014static} proposed a static approach for Android malware detection via extracting permissions and API usage. They confirm that the integration between the feature of permission and API calls can achieve a better precision than just the only feature of permission. Karbab et al.~\cite{karbab2017android} introduced an automatic and effective Android malware detection system, MalDozer, that depends on deep learning techniques and raw sequences of API method calls. Arp et al.~\cite{arp2014drebin} proposed a tool DREBIN that builds a SVM based detection model utilizing APIs and other related information. While Ma et al.~\cite{ma2019combination} de-compile the Android Apps and construct three different system API data sets: API usage, API frequency, and API sequence to detect malware. To be specific, Linares-V{\'a}squez et al.~\cite{linares2014mining} attempted to reveal the APIs and usage patterns related to energy consumption, and to provide potential guidance for developers to decrease energy consumption. 
\subsection{Special APIs}
Android APIs typically follow the general \emph{deprecated-replace-remove} evolution cycle. Work~\cite{li2020cda} introduced prototype tool called CDA to characterize deprecated APIs from different revisions. Their extensive investigation shows that the deprecated APIs are not continuously annotated and documented and over a half of these APIs are commented to provide alternatives but these alternatives are rarely replaced by the developers. Besides the aforementioned general publicly accessible APIs, there exists another type of API referred to as inaccessible API that can be recognized as internal or hidden. Internal APIs are resolved ones for system apps located in the package \texttt{com.android.internal} while hidden APIs are methods annotated by the javadoc \texttt{@hide}. Li et al.~\cite{li2016accessing} did an extensive investigation to reveal the usability of these APIs. They demonstrate that these inaccessible APIs are continuously implemented in the Android framework and used to access a specific set of features while without any promise of forward compatibility. They also reveal that there exist a plenty of apps are indeed calling these inaccessible APIs and the patterns of usage are quite different between each other.

\section{Conclusion}
\label{sec:conclusion}
In this paper, we presented an empirical study that investigates silently-evolved methods (or SEMs in short) in nine version pairs of the Android API. To perform the study, we built a prototype tool called \tool{}. This tool, given the Android framework codebase as an input,  is able to identify SEMs in the Android API. Using \tool{}, we found that SEMs are indeed present in the Android API, and a large number of them can be publicly accessed by app developers. Additionally, we empirically found that (i) the majority of publicly-accessible silently-evolved methods (or PASEMs in short) involve semantic changes and these changes could lead to field failures in their client apps, (ii) the majority of SEMs is only introduced into the framework once, indicating that framework maintainers may not have been aware of this situation, and (iii) PASEMs are frequently used by real-world Android apps, even without suitably using API version checks.

We foresee a number of venues for future work. First, we plan to run a user study in which we investigate how developers react to changes in the Android API and their related documentation. We believe that such a study would provide a better understanding on how to best document semantic changes in the Android API. Second, we plan to study how SEMs are related to bug reports by mining and analyzing issues on GitHub. Finally, we plan to investigate whether Android developers follow different development practices in minor and major version updates of the Android framework and how these practices relate to SEMs.

\section*{Acknowledgments}

This work was partially supported by the Australian Research Council (ARC) under a Laureate Fellowship project FL190100035, a Discovery Early Career Researcher Award (DECRA) project DE200100016, a Discovery project DP200100020, and a gift from Facebook.

\bibliographystyle{IEEEtran}
\bibliography{main}

\end{document}